\documentclass[twocolumn]{article}
\DeclareUnicodeCharacter{0308}{\"{}}
\usepackage[a4paper,top=2cm,bottom=2cm,left=1cm,right=1cm,marginparwidth=1.75cm]{geometry}
\usepackage{graphicx} 
\usepackage{amsmath}
\usepackage{comment}
\usepackage[numbers,sort&compress]{natbib}
\usepackage{color}

\usepackage{braket}
\usepackage[usenames, dvipsnames]{xcolor}
\usepackage{bm}
\usepackage{hyperref}
\hypersetup{
    colorlinks=true,
    linkcolor=blue,     
    urlcolor=blue,
    citecolor=blue
}

\usepackage{authblk}
\usepackage{marvosym}  
\usepackage[font=small]{caption}

\begin{document}

\title{Multiple Photon Field-induced Topological States in Bulk HgTe}

\author[1,2]{Dongbin Shin\thanks{{dshin@gist.ac.kr}}}
\author[2]{I-Te Lu}
\author[2]{Benshu Fan}
\author[2,3]{Emil Vi{\~n}as Bostr{\"o}m}
\author[2]{Hang Liu}
\author[4]{Mark Kamper Svendsen}
\author[2,5]{Simone Latini}
\author[2,6]{Peizhe Tang\thanks{{peizhet@buaa.edu.cn}}}
\author[2,3,7]{Angel Rubio\thanks{{angel.rubio@mpsd.mpg.de}}}

\affil[1]{Department of Physics and Photon Science, Gwangju Institute of Science and Technology (GIST), Gwangju 61005, Republic of Korea}
\affil[2]{Max Planck Institute for the Structure and Dynamics of Matter, Luruper Chaussee 149, 22761 Hamburg, Germany}
\affil[3]{Nano-Bio Spectroscopy Group, Departamento de F\'isica de Materiales, Universidad del Pa\'is Vasco, 20018 San Sebastian, Spain}
\affil[4]{NNF Quantum Computing Programme, Niels Bohr Institute,
University of Copenhagen, Universitetsparken 5, 2100 Copenhagen, Denmark}
\affil[5]{Department of Physics, Technical University of Denmark, 2800 Kgs. Lyngby, Denmark}
\affil[6]{School of Materials Science and Engineering, Beihang University, Beijing, China}
\affil[7]{Initiative for Computational Catalysis (ICC), The Flatiron Institute, 162 Fifth Avenue, New York, NY 10010, USA}

\maketitle

\section*{Abstract}
Strong light-matter interactions can be exploited to modify properties of quantum materials both in and out of thermal equilibrium.
Recent studies suggest electromagnetic fields in photonic structures can hybridize with condensed matter systems, resulting in photon field-dressed collective quantum states such as charge density waves, superconductivity, and ferroelectricity. 
Here, we show that photon fields in photonic structures, including optical cavities and waveguides, induce emergent topological phases in solids through polarization-mediated symmetry-breaking mechanisms.
Using state-of-the-art quantum electrodynamic density functional theory (QEDFT) calculations, we demonstrate that strong light-matter coupling can reconfigure both the electronic and ionic structures of HgTe, driving the system into Weyl, nodal-line, or topological insulator phases. 
These phases depend on the relative orientation of the sample in the photonic structures, as well as the coupling strength. 
Unlike previously reported laser-driven phenomena with ultrashort lifetimes, the photon field-induced symmetry breaking arises from steady-state photon-matter hybridization, enabling multiple robust topological states to emerge.
Our study demonstrates that vacuum fluctuations in photonic structures can be used to engineer material properties and realize rich topological phenomena in quantum materials on demand.

\section*{Introduction}
Light-matter interactions, induced by laser pulses and confined photon fields, have been highlighted as a tool for controlling the structural and electronic properties of quantum materials, resulting in the emergence of superconductivity, topological phase transition, and ferroelectric transition~\cite{basov_towards_2017,rudner_band_2020,bernevig_quantum_2006,Nova2019_ferro,Guan_WTe2_light_Weyl,de_la_torre_colloquium_2021,Jarc2023_TaS2_cavity,keren_cavity-altered_2025,I-Te_PNAS}.
Specifically, laser-induced metastable phenomena arise from diverse microscopic mechanisms, such as displacive excitations of coherent phonons via excited electronic structure, nonlinear phonon interactions, and Floquet dynamics~\cite{Sie2019_WTe2,shin_light-induced_2024,vaswani_light-driven_2020,mciver_light-induced_2020,disa_engineering_2021,zhou_pseudospin-selective_2023,bao_light-induced_2022,zhou_floquet_2023,fan_chiral_2024,bloch_strongly_2022,torre_colloquium_2021}.
For example, the Weyl semimetal (WSM) T$_d$-WTe$_2$ can be driven to undergo a topological phase transition into a trivial semimetal through a terahertz (THz) pump-induced shear distortion~\cite{Sie2019_WTe2}.
Additionally, a resonant THz pump can distort the lattice of the trivial semimetallic HgTe through nonlinear phononics, thereby triggering a transient topological phase transition into a WSM in its non-equilibrium~\cite{shin_light-induced_2024}.
These studies indicate that the topology of materials can be controlled by laser pumps, showing great potential for light-controlled quantum devices~\cite{basov_towards_2017,de_la_torre_colloquium_2021}.
However, these laser-induced phenomena face challenges in being utilized in optically controlled devices due to their non-equilibrium conditions characterized by short lifetimes ($\sim$ps) and heat dissipation~\cite{Sie2019_WTe2,shin_light-induced_2024,disa_engineering_2021,de_la_torre_colloquium_2021}.

\begin{figure}[t]
\includegraphics[width=0.95\linewidth]{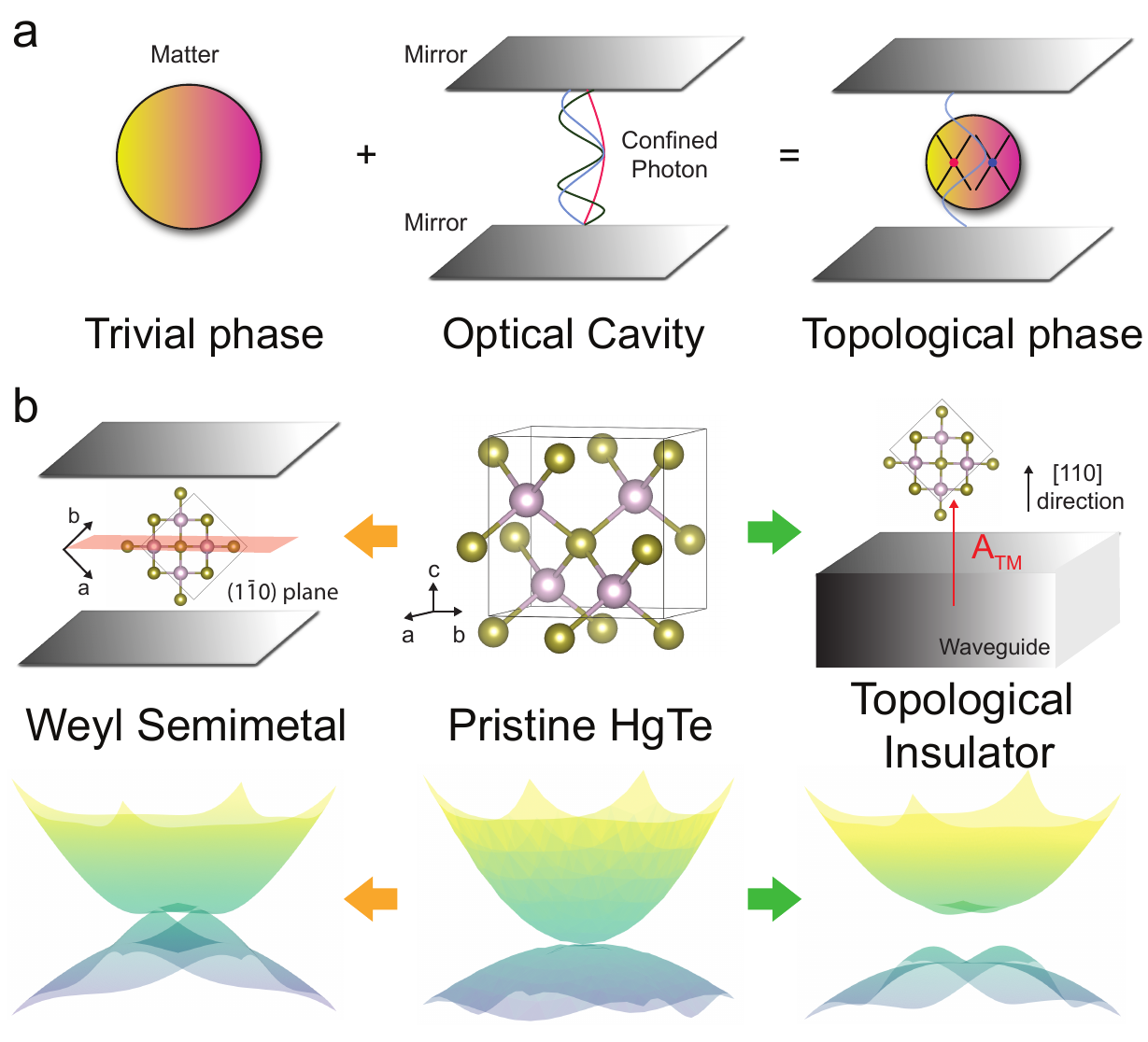}
\caption{
{\bf Photon field-induced topological phase transitions in HgTe via symmetry breaking.}
   \textbf{a}, Schematic image of strong coupling between matter and photons confined inside the optical cavity.
   \textbf{b}, Various topological phase transitions from trivial semimetal HgTe depending on the photon polarization directions with optical cavity and planar waveguide with ionic distortion. 
   In the bottom panels of (b), 2D Brillouin zone band structures along the $k_x-k_y$ plane with $k_x=k_y=0$ centered at $k_z=0$ are prepared at various topological phases.
}
\end{figure}

Photon fields, which are confined in photonic structures such as Fabry-Pérot cavities, plasmonic cavities, and planar waveguides, exhibiting quantum vacuum fluctuations, can modify material properties at equilibrium away from their pristine behaviors~\cite{Jarc2023_TaS2_cavity,Simone_PNAS,lu2025cavityengineeringsolidstatematerials,herzig_sheinfux_high-quality_2024,weber_cavity-renormalized_2023,dong_waveguide_2025,jin_reshaping_2018,stern_quantum_2024,hugall_plasmonic_2018,forn-diaz_ultrastrong_2019,Garcia_science2021}.
For instance, a Fabry-Pérot cavity, comprising two mirrors that confine vacuum-fluctuating photons (see Fig.~1a), has been used to modify the charge density wave transition temperature in bulk 1T-TaS$_2$ through strong coupling between the material and the vacuum photon field~\cite{Jarc2023_TaS2_cavity}. 
In this setup, the coupling strength can be tuned by adjusting the mirror separation or their tilting angle.
In addition, it is experimentally demonstrated that the vacuum fluctuating photon field can modify the integer and fractional quantum Hall effect~\cite{enkner_tunable_2025} and quench superconducting stiffness~\cite{keren_cavity-altered_2025}.
A recent theoretical study suggests that a confined photon field inside a Fabry-Pérot cavity can induce a ferroelectric ground state in the quantum paraelectric SrTiO$_3$ by forming a light–matter hybrid state. 
This process, driven by electromagnetic quantum fluctuations, is analogous to the mechanism of dynamical localization in driven systems~\cite{Simone_PNAS}.
Similarly, it has been theoretically proposed that such cavity-induced effects can enhance the critical temperature (T$_c$) of phonon-mediated superconductors like MgB$_2$~\cite{I-Te_PNAS}.
These examples highlight the potential of engineered photon fields to generate novel photon-matter hybrid states through strong light–matter interactions. 
In particular, photons can be tailored to interact with various degrees of freedom in condensed matter systems, including electrons, spins, orbitals, and phonons, leading to modifications in properties such as magnetism, superconductivity, and lattice vibrations.
Unlike the laser-induced phenomena out of equilibrium, photon fields in photonic structures modify material properties in the steady state of thermal equilibrium~\cite{Jarc2023_TaS2_cavity,Simone_PNAS,I-Te_PNAS,chiriaco_thermal_2024,pannir-sivajothi_blackbody_2025,schafer_shining_2022,svendsen_theory_2023,ruggenthaler_understanding_2023,sidler_perspective_2022,flick_atoms_2017}. 
In addition to operating in the quantum fluctuation regime, optical cavities can also be used in the driven steady-state or thermal modes (i.e., with a non-zero photon number), offering the advantage of not requiring strong external driving due to photon confinement~\cite{chiriaco_thermal_2024}.
Precise, stable, and long‐lived control over material properties is required to operate optically-controlled quantum devices, such as topological quantum computers, topological Josephson-junction circuits, and spin Hall effect devices~\cite{jungwirth_spin_2012,microsoft2025interferometric,nadj2014observation}. 
And material engineering via photon fields in photonic structures could provide a new approach to achieve this goal.

\begin{figure}[t]
\includegraphics[width=0.95\linewidth]{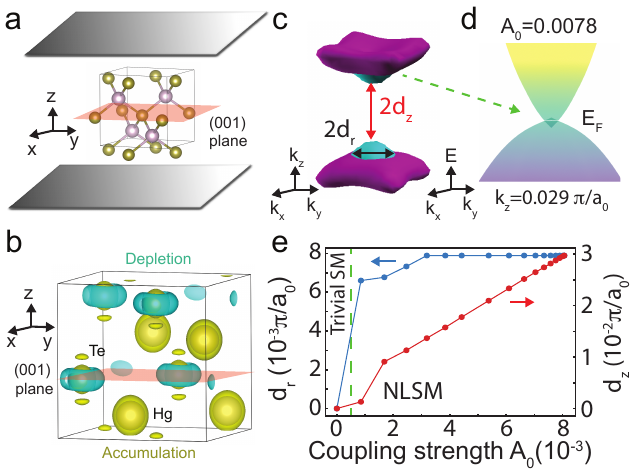}
\caption{
{\bf Topological nodal-line semimetallic phase in HgTe induced by two photons parallel with (001) plane.}
\textbf{a}, Schematic image of Fabry-Pérot cavity mirrors parallel with ($001$) plane of cubic HgTe structure.
\textbf{b}, The modified charge density $\Delta \rho(\vec{r})$ of HgTe by two photons parallel with the (001) plane. $\Delta \rho(\vec{r})=\rho_{\rm {QEDFT}}^{\textrm{A}_0=0.0078}(\vec{r})-\rho_{\rm {DFT}}(\vec{r})$, in which $\rho_{\rm {QEDFT}}^{\textrm{A}_0=0.0078}(\vec{r})$ and $\rho_{\rm {DFT}}(\vec{r})$ are charge densities under photon fields with the light-matter coupling strength $\textrm{A}_0=0.0078$ and without photon field.
\textbf{c-d}, Nodal lines in (c) Fermi surface plot and (d) 2D Brillouin zone band structures at $k_z=0.029 \pi/a_0$.
\textbf{e}, Evolution of the radius of nodal-line ($d_r$) and distances (2$d_z$) between nodal-lines with respect to light-matter coupling strength A$_0$.
Blue dots and line stand for the radius of nodal-line, and red dots and line stand for the distances between nodal-lines between Weyl points.
In (c), the purple and cyan surfaces indicate the hole and electron pockets.
}
\end{figure}

In this study, we explore cavity-mediated topological phase transitions in bulk HgTe using state-of-the-art QEDFT calculations~\cite{I-Te_PNAS,Christian_PNAS_QEDFT,ruggenthaler_quantum-electrodynamical_2014}. 
We evaluate the modified electronic and ionic structures induced by photon fields in various kinds of photonic structures (see Fig. 1b)~\cite{Jarc2023_TaS2_cavity,herzig_sheinfux_high-quality_2024,de_oliveira_graphene_2015,dong_waveguide_2025}.
The trivial semimetal HgTe can undergo a series of topological phase transitions to a nodal-line semimetal (NLSM), a WSM, and a topological insulator (TI), depending on the light-matter coupling strength and the relative orientation between the crystal structure of HgTe and the polarization direction of the photon field.
Unlike previous approaches for symmetry engineering in HgTe via external pressure~\cite{ruan_symmetry-protected_2016} or nonlinear phononics~\cite{shin_light-induced_2024}, by choosing photon fields in a suitable photonic structure, we could selectively engineer electronic structures of HgTe with or without ionic distortions to induce symmetry breaking and topological phase transitions without additional carrier pumping.
Importantly, our approach does not require resonance with the intrinsic photon frequency of the photonic structure. 
Thus, the predicted light-matter interaction effects are observable even under off-resonant conditions.
Our study suggests a potential approach to realize photon field-controlled topological states in a photon-matter hybridized system.

\begin{figure*}[h!]
\centering
\includegraphics[width=0.65\linewidth]{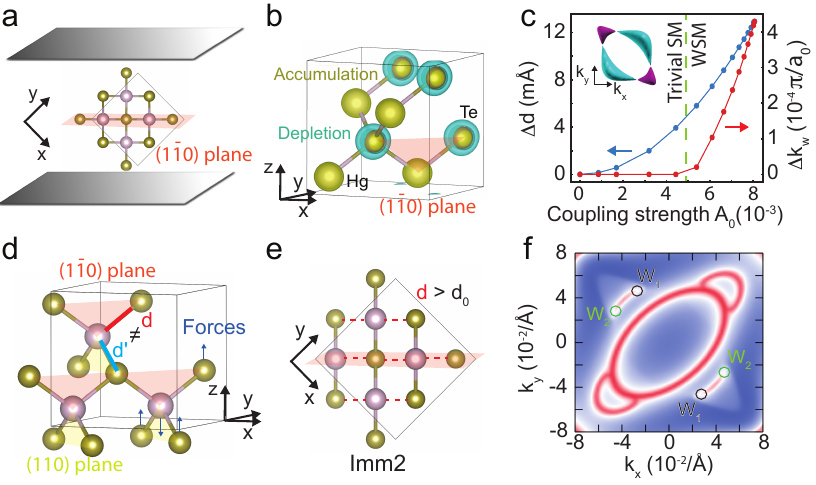}
\caption{
{\bf Weyl semimetal HgTe with two photons parallel with ($1\bar{1}0$) plane.}
\textbf{a}, Schematic image of Fabry-Pérot cavity mirrors parallel with ($1\bar{1}0$) plane of cubic HgTe structure.
\textbf{b}, The modified charge density of HgTe by two photons aligned ($1\bar{1}0$) plane.
\textbf{c}, Modified bonding distance ($\Delta d$) and Weyl points distance ($\Delta k_W$) with respect to light-matter coupling strength ($\rm {A_0}$). Herein, $\Delta d$ is defined as amplitude of atomic displacement from pristine geometry $\Delta d = |\vec{x}_\tau^{\textrm{A}_0}-\vec{x}_\tau^0|$.
\textbf{d-e}, The modified lattice distortion of HgTe by two photons aligned ($1\bar{1}0$) plane for (d) induced forces by modified charge density and (e) modified bonding distance.
\textbf{f}, Fermi arc surface state achieved by cavity-induced Weyl points with $w_1=0.049 \pi/a_0$ and $w_2= -0.023\pi/a_0$ values under the coupling strength A$_0=0.0078$.
In (a), (b), (d), and (e), orange planes indicate the ($1\bar{1}0$) plane. 
The inset of (c) indicates the Fermi surface, while the purple and cyan surfaces indicate the hole and electron pockets.
In (d), the yellow plane stands for the ($110$) plane, and the blue arrows indicate induced forces along the z-direction by the modified charge density.
In (e), the red dashed lines indicate the modified bonding distance between Hg and Te atoms.
}
\end{figure*}

\section*{Results}
The pristine HgTe, in a zinc-blende structure, is a topologically trivial semimetal with a four-fold degenerate point between parabolic valence and conduction bands at the Fermi level (see middle panel of Fig.~1b).
Constrained by its lattice symmetry, bulk HgTe belongs to the F$\bar 4$3m point group and allows 24 symmetry operations.
Upon external perturbation, emergent topological phases are unraveled in HgTe~\cite{shin_light-induced_2024,bernevig_quantum_2006,konig_quantum_2007,ruan_symmetry-protected_2016}.
For example, a CdTe/HgTe/CdTe quantum well structure with a sufficient thickness of HgTe was proposed and realized as the first quantum spin Hall insulator~\cite{bernevig_quantum_2006,konig_quantum_2007}.
External pressure, by altering the intrinsic lattice of three-dimensional bulk HgTe, can induce topological phase transitions, resulting in either a WSM or a TI phase, depending on the direction of the applied pressure~\cite{ruan_symmetry-protected_2016}. 
These studies underscore the pivotal role of symmetry breaking in driving topological transitions in HgTe. 
In this work, we aim to achieve such symmetry breaking through a photon field embedded in a photonic structure. 
Using extensive QEDFT calculations, we propose a mechanism by which photon fields can induce topological phase transitions in HgTe.

We first consider HgTe coupled to the two polarizations of the fundamental cavity mode of a Fabry-Pérot type resonator, such that the electric field is parallel with the ($001$) plane of the HgTe cubic lattice (see Fig.~2a).
With a given light-matter coupling strength ($\rm {A_0}=\tilde{\lambda}/\sqrt{2\tilde{\omega}}=0.0078$ in atomic Hartree unit), defined as the ratio between the photon-matter interaction coefficient ($\tilde{\lambda}$ in energy) and the frequency ($\tilde{\omega}$ in time$^{-1}$) of dressed photon~\cite{I-Te_PRA_QEDFT}, we evaluate the photon induced modification of the electronic structure and resulting atomic forces on each atom via the QEDFT approach [see Method part and Supplementary Information (SI) for details].
In this configuration, photon fields change the electron charge density compared to its intrinsic ground state, leading to the formation of a new light-matter hybrid state. 
From the charge density difference $\Delta \rho(\vec{r})$ shown in Fig.~2b, we can observe the charge depletion around the ($001$) plane for Te atoms (green isosurface) and spherical charge accumulation for Hg atoms (yellow isosurface), indicating that photon fields parallel with the ($001$) plane can introduce an anisotropic charge redistribution, breaking the intrinsic crystal symmetry of HgTe. 
At the same time, our QEDFT calculation shows no net forces on each ion with this charge modification, indicating no change in the atomic structure of the HgTe in such a photonic structure. 
Figures 2c and 2d show the Fermi surface and calculated band structures of bulk HgTe modified by the photon fields. Instead of a four-fold degenerate point at the Fermi level, characteristic of intrinsic HgTe, two gapless nodal-lines can be observed in the $k_x-k_y$ plane with different momentum ($\pm d_z$) along the $k_z$ direction. 
This suggests a topological phase transition, which is only contributed by the cavity-induced charge density modification and no ionic distortion. Notably, the values of the distance ($d_z$) and the radius ($d_r$) of nodal-lines can be controlled via the light-matter coupling strength $\rm {A_0}$. 
As $\rm {A_0}$ is enhanced, $d_z$ and $d_r$ increase correspondingly (see Fig. 2e).
This behavior indicates that bulk HgTe inside an optical cavity, such that the ($001$) plane is parallel with the mirrors, becomes an NLSM, and its topological features can be efficiently modulated by the light-matter coupling strength.

\begin{figure*}[t]
\centering
\includegraphics[width=0.65\linewidth]{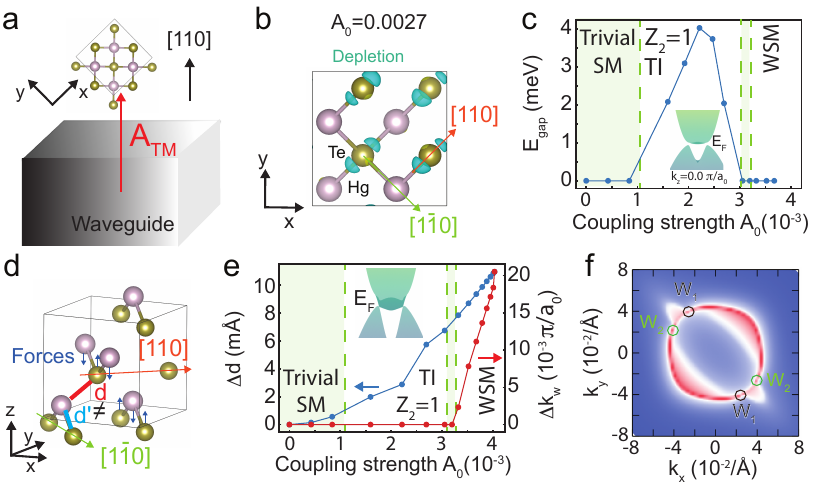}
\caption{
{\bf Three-step topological phase transitions by a single photon polarized along the [$110$] direction.}
\textbf{a}, Schematic image of HgTe geometry with a single photon along the [$110$] direction induced by the waveguide's TM mode.
\textbf{b}, The modified charge density of HgTe by a single photon polarized along the [$110$] direction.
\textbf{c}, Modified band gap ($\rm{E_{gap}}$) with respect to cavity coupling strength ($\rm {A_0}$). These regions with a green background stand for the semimetallic phase. The region with lower coupling strength (A$_0$ $\textless 0.0012$) is topologically trivial, and the $Z_2$ number is 1 for the phase with larger coupling strength ($0.0012$~ $\leq$ A$_0$ $\leq 0.0027$). 
\textbf{d}, Modified lattice distortion of HgTe by a single photon polarized along the [$110$] direction.
\textbf{e}, Modified bonding distance ($\Delta d$, blue) and Weyl points distance ($\Delta k_W$, red) with respect to coupling strength ($\rm {A_0}$). Blue dots and line stand for bonding distance, and red dots and line stand for the distance between Weyl points.
\textbf{f}, Fermi arc surface state on ($001$) side surface for the 1D photon field-driven HgTe with $\rm {A_0}=0.0039$. Hollow circles stand for four Weyl points.
In (a), (b), and (d), red arrows indicate [$110$] direction.
In (c) and (e), the green boxes indicate the trivial semimetallic phase region.
Insets of (c) and (e), 2D Brillouin zone band structure at $k_z=0.0$ with $\rm {A_0}=0.0027$ and $\rm {A_0}=0.0039$, respectively.
In (d), a green arrow indicates [$1\bar{1}0$] direction, and the blue arrows indicate induced forces along the z-direction by the modified charge density.
}
\end{figure*}

The topological behavior of HgTe can be further engineered by tuning the lattice orientation within the optical cavity, enabling simultaneous modifications to both ionic and electronic degrees of freedom. 
This allows the realization of a WSM phase in cavity-driven HgTe.
Unlike the previous configuration, here we rotate the HgTe crystal such that the ($1\bar{1}0$) plane is aligned parallel to the cavity mirrors (see Fig.~3a).
In this configuration, the light-matter coupling also modifies the charge density (see Fig.~3b), resulting in a charge depletion at the Te atomic site on the ($1\bar{1}0$) plane and a spherical charge accumulation at the Hg atomic sites along the photon field direction. 
In contrast to the previous geometry, the charge redistribution on the Te atomic sites along ($1\bar{1}0$) plane introduces asymmetric interactions between upper Hg-Te bonds in the ($1\bar{1}0$) plane and bottom Hg-Te bonds in the ($110$) plane (see Fig.~3d).
This asymmetry results in a non-zero net force on Hg and Te atoms and induces an ionic distortion ($\Delta d$) along the z direction correspondingly (see the blue arrows in Fig.~3d). 
Our QEDFT calculations confirm such observations and find that the space group of the relaxed HgTe structure becomes $Imm2$ (see Fig. 3e) with a lattice distortion, the value of which is directly proportional to the light-matter coupling strength (see blue line in Fig. 3c).

The modifications of the electronic and ionic structure discussed above can lead to the emergence of Weyl points around the Fermi level (see Fig. 3f). As shown in Fig. 3c, with increasing the light-matter coupling strength, the system hosts a trivial semimetal phase at the low strength limit, then becomes a WSM when $\rm {A_0} \ge 0.005$. 
In the latter phase, the distance between two Weyl points ($\Delta k_W=|k_{W_1}-k_{W_2}|$) increases with light-matter coupling strength as the ionic distortions increase (see Fig. 3c).
Four Weyl points stay around the Fermi level within $k_z=0$ plane as $(w_1,w_2,0)$, $(-w_1,w_2,0)$, $(w_2,w_1,0)$, and $(-w_2,-w_1,0)$, and the Fermi arc states that connect them on the side surface are shown in Fig.~3f. 

To distinguish electronic and ionic contributions, we perform QEDFT calculations where we artificially fix atomic positions to those within the F$\bar 4$3m point group, and take only the charge redistribution induced by the photon fields into account. HgTe in such a fixed geometry remains a trivial semimetal with the same photon field strength. 
This result indicates that the WSM phase in HgTe inside the cavity, with the ($1\bar{1}0$) plane parallel to the cavity mirrors, is truly non-perturbative as it originates from the lattice symmetry breaking and consequent structural rearrangement.
The SI provides details on the electronic structure analysis for the various conditions, i.e. fixed geometry and light-matter coupling strength, for the case of two photon modes parallel with the ($1\bar{1}0$) plane.

\begin{table*}[t]
\centering
\caption{Cavity-induced topological phase transition from trivial semimetallic phase and modified space group from F$\bar{4}$3m depending on photon polarization direction}\label{tab1}%
\begin{tabular}{c|c|c|c|c|c}
\textrm{ photon directions} & [110]  & [111] & (1$\bar{1}$0) &  (111) &  (001) \\
\hline
\# of ph. 
polarizations  &  1 & 1 &  2 & 2 & 2  \\
\hline
1st phase  &  TI  & TI &  WSM & WSM & NLSM  \\
\hline
2nd phase  &  Semimetal &  Semimetal  &    &   &   \\
\hline
3rd phase  &  WSM &    &    &   &   \\
\hline
Space group & Imm2 &  Cm(8) &   Imm2  &  Cm(8) & F$\bar{4}$3m
\end{tabular}
\end{table*}

As a final configuration, we consider HgTe coupled to a single photonic mode with out-of-plane polarization aligned along the [$110$] direction of the crystal (see Fig.~4a). 
Such a mode can be hosted, for example, by a dielectric planar waveguide slab as a TM mode or as a hyperbolic phonon-polariton mode. 
The latter, recently demonstrated in hexagonal boron nitride, enables extreme sub-diffraction electromagnetic confinement and has even been shown to alter superconductivity~\cite{de_oliveira_graphene_2015,herzig_sheinfux_high-quality_2024,keren_cavity-altered_2025}. 
As the light-matter coupling strength increases in this out-of-plane setup, we observe multiple topological phases, including TI and WSM. 
Similar to the Fabry-Pérot cavity case, the photon field modifies the charge distribution, with noticeable depletion around Te atoms along the polarization direction (see Fig.~4b).
Due to the redistribution of the charge density and minor ionic displacement, HgTe becomes a TI with a topological number of $Z_2=1$ and a band gap opening around the $\Gamma$ point, when $0.0012$~$\leq$ A$_0$ $\leq 0.0027$~(see Fig.~4c).
For a coupling strength in the region of $0.0027$~$\textless$ A$_0$ $\textless$~$0.0033$, a semimetal phase can be found, in which electronic states around the $\Gamma$ point maintain topologically non-trivial properties while both electron pockets and hole pockets staying at different $k$ points can be observed at the Fermi level (see Fig.~S4 in SI).

Upon further increase of the light-matter coupling strength (A$_0 \geq 0.0033$), the contribution from lattice distortion becomes prominent and contributes further to the change of electronic structures. Under strong out-of-plane phonon fields, the charge depletion for Te atoms along [$110$] direction leads to asymmetric bonding configurations for the two Hg-Te bonds aligned with the [$110$] direction, resulting in a net force along the [$001$] direction and reducing the lattice symmetry to the Cm(8) space group (see Fig.~4d). 
The consequent large lattice distortion leads to the formation of four Weyl points around the Fermi level (see Figs.~4d and 4e), whose projections on the ($001$) surface are connected by the Fermi arc states shown in Fig.~4f.
We note that, in same light-matter coupling strength window  (with A$_0 \geq 0.0033$), if we do not allow for structural relaxation keeping the pristine lattice structure (F$\bar 4$3m point group), the electronic charge modification alone does not alter the topological phase of HgTe (as TI) and only increases the band gap (see Fig.~S3 in SI).
This result indicates that the self-consistent cooperation between charge and lattice modifications, induced by a single photon field polarized along the [$110$] direction, is necessary to achieve the complex topological phase diagram in bulk HgTe.
In the SI, we discuss further details of electronic structure analysis under various conditions, such as fixed geometry and different light-matter coupling strengths.

These examples demonstrate how varying the orientation of bulk HgTe, strongly coupled to different photon modes, can produce a rich variety of complex topological phases. 
Table 1 summarizes multiple topological phases observed in bulk HgTe under different photon fields and crystal alignments. 
We found that two-photon modes aligned along the ($111$) plane induce Weyl points through ionic distortions (see Fig.~S2 in the SI). 
Additionally, a one-photon mode oriented along the [$111$] direction drives a two-step topological phase transition from a TI to a semimetal with $Z_2=1$ (see Fig.~S5 in the SI). 
These findings highlight that the photonic environment can induce symmetry breaking in materials by modifying both electronic and ionic structures, enabling the emergence of multiple topological phases in bulk HgTe.

\section*{Discussion}
Vacuum photon fluctuations induced topological phase transitions are studied using state-of-the-art QEDFT calculations.
We show that topological phases of bulk HgTe can be selectively controlled depending on the orientations of bulk HgTe relative to the polarization of photonic modes of an optical cavity and a waveguide of dielectric or hyperbolic type.
The photon-matter interaction induces changes of the electronic charge density, in a process similar to the dynamical localization in driven systems~\cite{Simone_PNAS}, which in turn makes it possible to introduce ionic distortion and crystal symmetry breaking.
As a result, the emergent topological phases in photon field-dressed HgTe can be selectively activated, depending on the photon field polarization directions and strengths.
When the ($001$) plane of HgTe is parallel to the two polarizations of the mode of an optical cavity, the trivial semimetal HgTe becomes an NLSM via photon field-induced electronic charge rearrangement but without crystal structure changes.
When, instead, the (1$\bar{1}$0) plane of HgTe is parallel with the two in-plane photon fields of the optical cavity, the topological phase transition to WSM follows from ionic distortion and related crystal symmetry breaking.
Finally, when a single photon mode of a dielectric planar waveguide is coupled to HgTe along the [$110$] direction, the system displays a TI phase within a weak photon field-matter coupling regime, and a WSM phase under strong coupling, as a consequence of ionic distortions.

The range of light-matter coupling strength ($0\leq \rm {A_0} \leq 0.008$) explored in our work is realistically achievable in hyperbolic photon-polariton or plasmon-polariton based photonic structures in their ground states, even in the absence of real photons (dark cavity)~\cite{I-Te_PNAS,jin_reshaping_2018,stern_quantum_2024}.
On the other hand, such strong couplings for the two-mirrored optical cavity configuration remain challenging to achieve by photon vacuum‐fluctuation only. 
However, in principle, they can be reachable in the case of a non-zero photon population, such as a resonantly pumped Fabry-Pérot cavity~\cite{kim_fabryperot_2023,li_fabry-perot_2025,huang_fabry-perot_2019,zhou_cavity_2024,grunwald_cavity_2025}.
In the SI, the light-matter coupling strengths ($\rm {A_0}$) with vacuum fluctuating photons and pumped photons are discussed.
Furthermore, additional future studies that bridge theoretical models and experimental observations will be helpful to fully understand the feasibility of strong light-matter interaction of vacuum-fluctuating photon fields in different types of photonic environments~\cite{Jarc2023_TaS2_cavity,chiriaco_thermal_2024,pannir-sivajothi_blackbody_2025,fassioli_controlling_2024}. 

Photon field-induced phenomena present a clear advantage over strain-induced or laser-driven approaches. 
By selecting the appropriate sample orientation and tuning the coupling strength within a photonic structure, multiple emergent topological states in bulk HgTe can be accessed and controlled under ambient conditions. 
In contrast, pressure-induced transitions require high or anisotropic pressures to drive HgTe into a topological phase~\cite{ruan_symmetry-protected_2016}, while laser-driven methods~\cite{shin_light-induced_2024} produce only short-lived, non-equilibrium states that quickly succumb to irreversible heating, limiting their potential for device applications. 
Our study suggests that photon fields confined within optical cavities or waveguides offer a more robust and versatile platform for manipulating topology, with both geometry and coupling strength serving as effective tuning parameters.

\section*{Method}
We used the custom-made QEDFT implementation within the Quantum Espresso package to evaluate the electronic structure, the ionic relaxation, and photon-matter interactions~\cite{Giannozzi2017,I-Te_PRA_QEDFT,I-Te_PNAS}.
A plane-wave basis up to $80$~Ry was employed to describe the Kohn-Sham wavefunctions, and spin-orbit coupling was included.
The Perdew-Wang-type local density approximation functional is considered to describe the electron-electron exchange and correlation effects~\cite{perdew_wang}.
The core-level states were considered via the projector augmented wave method.
We sampled the Brillouin zone with a $6 \times 6 \times 6$ grid of points for the HgTe cubic superlattice with 4 Hg and 4 Te atoms and checked the k-point convergence up to $9 \times 9 \times 9$ and the consistency in its primitive lattice.
To investigate the topological properties of HgTe with the Wannier interpolation technique, we used Wannier90 and WannierTools packages~\cite{pizzi_wannier90_2020,wu_wanniertools_2018}.
Bloch states with $200 \times 200 \times 200$  and $200 \times 200$ sampled grid points in the Brillouin zone were considered for finding the Fermi level and the Fermi arc surface plot, respectively.

To study the effect of photon fields on electronic and ionic structures from first principles, we use the QEDFT framework based on the non-relativistic Pauli-Fierz Hamiltonian~\cite{I-Te_PRA_QEDFT}. 
The ground state properties of the electronic subsystem can be obtained via the auxiliary Kohn–Sham (KS) scheme. The KS Hamiltonian reads

\begin{equation}
\hat{H}_{\rm KS} = -\frac{\hbar^2}{2m_e} \nabla^2 + v_{\rm KS}(\mathbf{r}),
\end{equation}
where the KS potential $v_{\rm KS}(\mathbf{r})$ comprises the external potential $v_{\rm ext}(\mathbf{r})$, the Hartree and electronic exchange–correlation potential $v_{\rm Hxc}(\mathbf{r})$, and the photon–electron exchange–correlation potential $v_{\rm pxc}(\mathbf{r})$.
For the latter, we adopt the electron-photon exchange approximation within the local density approximation (pxLDA)~\cite{I-Te_PRA_QEDFT}. The corresponding pxLDA potential is determined by solving the Poisson equation:
\begin{equation}
\nabla^2 v_{\rm pxLDA}(\mathbf{r}) = -\sum_{\alpha=1}^{M_p} \frac{2\pi^2 \tilde{\lambda}^2_\alpha}{\tilde{\omega}^2_\alpha} \left[ (\tilde{\boldsymbol{\varepsilon}}_\alpha \cdot \nabla)^2 \left( \frac{3\rho(\mathbf{r})}{8\pi} \right)^{\frac{2}{3}} \right].
\end{equation}

Here $\tilde{\lambda}_\alpha$, $\tilde{\omega}_\alpha$, and $\tilde{\epsilon}_\alpha$ are the light-matter interaction coefficient, frequency,
and polarization for the $\alpha$-th dressed photon mode. 
The connection between dressed and bare photon frequencies can be found in Refs.~\cite{I-Te_PNAS} and~\cite{I-Te_PRA_QEDFT}.
Although a specific photon frequency $\tilde{\omega}_{\alpha}$ is used, the strength of the photon field effect in the pxLDA approximation is largely governed by the ratio $\tilde{\lambda}_{\alpha} / \tilde{\omega}_{\alpha}$, if we consider one or two orthogonal effective photon modes with identical frequency and coupling strength. 
In our work, we denote the light–matter coupling strength $\tilde{\lambda}_{\alpha}/ \sqrt{2 \tilde{\omega}_{\alpha}}$ as $\rm {A_0}$, which is comparable with the amplitude of the vector potential of photon field in the non-relativistic Pauli-Fierz Hamiltonian.
Finally, the forces acting on nuclei are evaluated using the Hellmann–Feynman theorem.

\section*{Acknowledgments}
We acknowledge support by the Max Planck Institute New York City Center for Non-Equilibrium Quantum Phenomena, the Cluster of Excellence “CUI: Advanced Imaging of Matter”–EXC 2056–Project ID 390715994, European Research Council (ERC-2024-SyG-101167294; UnMySt), and Grupos Consolidados (IT1453-22).
P.T. was supported by the National Natural Science Foundation of China (Grants No. 12234011 and No. 12374053) and the National Key Basic Research and Development Program of China (Grant No. 2024YFA1409100).
D.S. was supported by the National Research Foundation of Korea (NRF) grant funded by the Korea government (MSIT) (No. RS-2024-00333664) and the Ministry of Science and ICT (No. RS-2022-NR068223).
M.K.S. is supported by the Novo Nordisk Foundation, Grant number NNF22SA0081175, NNF Quantum Computing Programme. 
We also acknowledge support from the Max Planck–New York Center for Non-Equilibrium Quantum Phenomena.
The Flatiron Institute is a division of the Simons Foundation.

\section*{Author Contributions}
 D.S., S.L., P.T., and A.R. wrote the manuscript. 
 D.S. performed the ab initio calculations under the supervision of P.T. and A.R.. I.L. and A.R. developed the QEDFT code. 
 D.S., B.F., and P.T. analyzed the topological phase of HgTe. E.V.B., M.K.S., I.L., and S.L. investigated the condition of realistic light-matter coupling strength. 
 All authors discussed the results and contributed to the final paper.
 
\end{document}